\documentclass[aps,superscriptaddress,reprint,longbibliography,onecolumn]{revtex4-2}
\usepackage{amsmath}
\usepackage{graphicx,textcomp}
\usepackage{enumerate,subeqnarray}
\usepackage{verbatim,natbib}
\usepackage{amsmath, amsthm, amssymb}
\usepackage{amsfonts}
\usepackage[colorlinks=true, citecolor=red, linkcolor=blue, urlcolor=blue ]{hyperref}

\newcommand{\oo}[1]{\mathcal O}
\newcommand{\xhat}{\hat{\mathbf x}}
\newcommand{\yhat}{\hat{\mathbf y}}
\newcommand{\zhat}{\hat{\mathbf z}}
\newcommand{\eninv}{\epsilon_0^{-1}}
\newcommand{\lm}{\Lambda_-}

\newcommand{\lp}{\Lambda_+}

\usepackage{orcidlink}
\begin{document}
	
	\title{Load-dependent resistive-force theory for helical filaments}
	
	\author{Pyae Hein Htet\,\orcidlink{0000-0001-5068-9828}}
	\affiliation{Department of Applied Mathematics and Theoretical Physics, University of Cambridge,	
		Cambridge CB3 0WA, UK}
	\author{Eric Lauga\,\orcidlink{0000-0002-8916-2545}}
	\email{e.lauga@damtp.cam.ac.uk}
	\affiliation{Department of Applied Mathematics and Theoretical Physics, University of Cambridge,	
		Cambridge CB3 0WA, UK}
	
	\date{\today}

	\begin{abstract}
		
		The passive rotation of rigid helical filaments is the propulsion strategy used by flagellated bacteria and some artificial microswimmers to navigate at low Reynolds numbers.  In a classical 1976 paper, Lighthill calculated the `optimal' resistance coefficients in a local (logarithmically accurate) resistive-force theory that best approximates predictions from the nonlocal (algebraically accurate) slender-body theory for force-free swimming of a rotating helix without an attached load (e.g., no cell body).   These coefficients have since been widely applied, often beyond the conditions for which they were originally derived.  
	Here, we revisit the problem for the case where a load is attached to the rotating filament, such as the cell body of a bacterium or the head of an artificial swimmer.    We show that the optimal resistance coefficients depend in fact on the size of the load, and we quantify the increasing inaccuracy of Lighthill's coefficients as the load grows. Finally, we provide a physical explanation for the origin of this unexpected load-dependence.

	\end{abstract}

	\maketitle

	\section{Introduction}
	
	Slender appendages are ubiquitous in cellular biology and play a crucial role in locomotion and transport, from large cilia and flagellated spermatozoa   to the much smaller bacteria~\cite{lauga2020fluid}. In particular, the passive rotation of rigid helical filaments provides a means of overcoming the Stokes-reversibility constraints known in the context of locomotion as the scallop theorem~\cite{purcell1977}. Indeed this is the strategy used by many flagellated bacteria, such as the model organism \textit{E.~coli}, to self-propel at low Reynolds numbers~\cite{wadhwa2022, berg2004, lauga2016}. Spirochetes are an important family of bacteria which self-propel by exerting forces along their helical body using `endoflagella' between the inner and outer cell membranes~\cite{vig2012,li2000},  {and the unique biology of these endoflagellar systems is a topic of active research \cite{sanmartin2023}. Helical flagella have biological implications beyond locomotion, as reported in recent work highlighting a correlation between flagellar rotational direction and biofilm formation in \textit{Helicobacter pylori} \cite{liu2024}. \textit{Bacillus subtilis} bacteria navigate anisotropic media by organising their numerous flagella at different sides of their cell body and activating/deactivating these flagella \cite{prabhune2024}.}
	
	These helical biological swimmers have, in turn, inspired the design of field-driven helical micro/nanorobots and `artificial bacterial flagella'~\cite{zhang2009, ghosh2018} with promising potential for biomedical applications 
	such as chemical delivery and  microsurgery~\cite{qiu2015}.  {A recent review offers a comprehensive discussion of the latest developments in helical micro/nanoswimmers, addressing their actuation mechanisms, fabrication methods, functionalisation strategies, and vast clinical applications, as well as the challenges in transitioning from laboratory research to \textit{in vivo} applications \cite{dong2022}. For example, a biohybrid helical nanoswimmer made of gold nanoparticles integrated into {\it Spirulina} algae has been developed, demonstrating outstanding capabilities for imaging and targeted therapy \cite{li2024}. On the engineering side, novel strategies have been proposed to enhance their swimming performance in viscous environments, including nested helical tails \cite{zhang2023}, ribbon-like tail geometries \cite{piranfar2024}, and modification of surface microstructure \cite{wang2024}. The fluid dynamics of helical filaments is therefore an important, and active, field of research, from biological, biophysical, and engineering perspectives.}

	
	The fundamental physical reason behind the propulsion generated by undulating slender filament    is the anisotropic drag exerted on it by a viscous fluid, which allows the generation of a thrust force in a direction normal to the filament undulation or rotation~\cite{lauga2009, hancock1953, becker2003}. 	The exact Stokes flow solution around a translating prolate spheroid~\cite{chwang1976} confirms the intuitive idea that a filament experiences more drag, for the same speed, when it translates in a direction normal to the filament centreline than when it translates in the parallel direction; the  ratio of drag becomes  exactly two in the asymptotic limit of very slender shapes.

	
	Biological slender filaments such as eukaryotic flagella and bacterial flagellar filaments are often curved, and in these cases the flow generated can be quantified by asymptotic techniques which mathematically exploit their  slenderness. 	
	Perhaps the simplest approach to understanding the fluid dynamics of a curved filament is what is now known as resistive-force theory (RFT). First proposed in  seminal work by Gray and Hancock~\cite{hancock1953}, this approach derives  an anisotropic proportionality relationship, at every point along the filament, between the local velocity of the filament and the force it locally exerts onto the ambient fluid. 	There is some debate as to the appropriate values of these proportionality constants in this framework, and later studies have derived more accurate values of these coefficients~\cite{cox1970}, including coefficients for filaments of arbitrarily shaped cross-sections~\cite{batchelor1970}.  		RFT, in essence,  {neglects hydrodynamic interactions from distant sections of the filament, and thus it is only logarithmically correct in the filament aspect ratio, but its simplicity makes it very well suited to gaining intuitive understanding and qualitative insight into biophysical systems.}

	A more quantitatively accurate framework to describe the hydrodynamics of a curved filament is known as slender-body theory (SBT), which accounts for contributions to the flow from the entire filament by placing an appropriate distribution of flow singularities along the filament centreline.  	This yields a nonlocal integral relation between the force distribution and the filament velocity, algebraically correct in the filament aspect ratio~\cite{lighthill1975mathematical}.  {More refined SBTs 
		have since been derived, for instance those} which allow for twisting or dilating filaments~\cite{keller1976}, or account accurately for end effects and determine the force distribution up to second order in the filament aspect ratio~\cite{johnson1980}.   {More recent work presents SBTs for other filament cross-sections, such as ribbons \cite{koens2016} and arbitrary cross-sections \cite{borker2019}.} 	Although SBT is algebraically accurate, this comes at the cost of requiring the inversion of an integral relation, typically numerically~\cite{Gotz2000}.

	In the case of helical filaments, however, it is possible to achieve the accuracy of a slender-body theory within a much simpler RFT framework. This was demonstrated by 
	Lighthill in a classical 1976 paper~\cite{Lighthill1996HelicalDO}, in which he constructed an 'optimal' RFT with resistance coefficients modified to account for hydrodynamic interactions along the entire filament, in the specific case of the force-free swimming of a torqued helix without a load (i.e.~an isolated helix).  {By exploiting helical symmetry, Lighthill solved the SBT integral equations analytically, and then determined the RFT coefficients which best matched these SBT results~\cite{Lighthill1996HelicalDO}.} 
	These resistance coefficients, however, have since been used ubiquitously in the literature even when the conditions under which they were originally derived do not hold. Using Lighthill's coefficients in the presence of a load (e.g.~a bacterial cell body), is not only mathematically ill-founded, but also leads to significant quantitative inaccuracies.

	In this study we revisit the problem of deriving optimal resistance coefficients for a slender helical filament, in the case where an external load is attached, thereby improving upon the mathematical inconsistency and physical inaccuracies resulting from using Lighthill's coefficients outside their regime of validity.  We consider two  model problems as relevant to both biological and artificial swimmers:  {the force-free swimming of a torqued helical filament attached to an external load and the force- and torque-free swimming of a helical swimmer}.  We solve both problems   	using SBT and RFT, and by matching the results from these two approaches, determine the optimal RFT coefficients which reproduce the SBT results in both problems. 	We arrive at the surprising revelation that these optimal coefficients depend on the size of the load, and we further provide a physical explanation for this load-dependence in the context of hydrodynamic interactions between different parts of the filament.

	The paper is organised as follows.  
	After our problem statement (\S\ref{sec:problemsetup}), we derive a general semi-analytical solution using SBT and validate our solution against a numerical implementation of SBT (\S\ref{sec:generalsolution}). We then  {use this general solution to obtain the SBT results for the swimming speeds of a force-free  helical swimmer subject to an external torque, relevant to artificial swimmers(\S\ref{sec:swimmingproblems}\ref{sec:forcefree}), and of a force- and torque-free helical swimmer}, relevant to bacteria (\S\ref{sec:swimmingproblems}\ref{sec:torquefree}). We next investigate the inaccuracies resulting from using Lighthill's coefficients (\S\ref{sec:rft}). 
	Finally we derive the new load-dependent resistance coefficients which remedy these inaccuracies, and offer a physical explanation for the new result of an apparent load-dependence for the optimal resistance coefficients (\S\ref{sec:optimalcoeff}).

	\section{Problem setup}\label{sec:problemsetup}

\begin{figure}[t]\centering
		\includegraphics[width=0.99\textwidth]{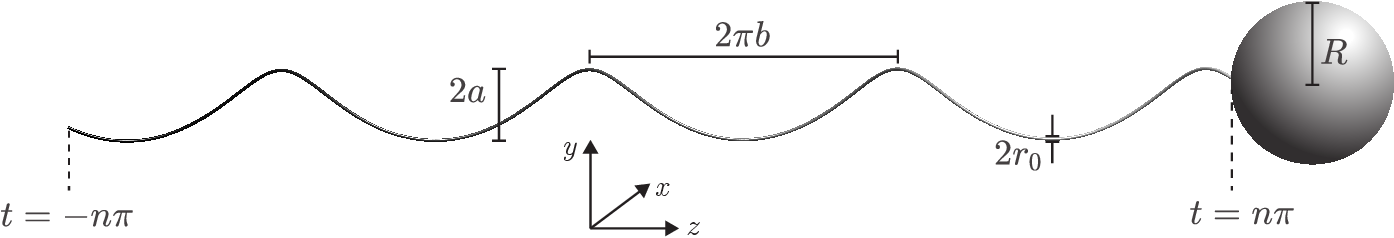}
		\caption{{
				\bf Viscous propulsion by helical filament}. A helical filament of helical radius $a$, pitch $2\pi b$, filament radius $r_0$ and $n$ turns (where $n$ is not necessarily an integer), coupled to a spherical load of radius $R$. The filament centreline is parametrised by $t \in [-n\pi,n\pi]$. {The shape illustrated has geometrical parameters $\epsilon = 0.61, \epsilon_0 = 0.028, n = 3.8$ corresponding to an \textit{E.~coli} normal polymorph (see Table~\ref{table}).} }\label{fig:setup}
	\end{figure}

	We consider the  model problem illustrated in Fig.~\ref{fig:setup}. We consider a right-handed helical filament of helical radius $a$, pitch $2\pi b$, and filament radius $r_0$, with  $n$ turns,  {where $n$ is not necessarily an integer.}  The basis vectors $\{\xhat, \yhat, \zhat\}$ are defined such that the centreline of the helix is parametrised by
	\begin{equation}
		\mathbf r(t) = (a \cos t, a \sin t, bt), -n\pi \leq t \leq n\pi,
	\end{equation} 
	where the parameter $t$ is related to the arclength $s$ by $\mathrm ds = \sqrt{a^2 + b^2} \mathrm dt$. We define the dimensionless geometrical parameters $\epsilon$ and $\epsilon_0$ as
	\begin{equation}
		\epsilon = \frac{a}{b}, \qquad \epsilon_0 = \frac{r_0}{{b}}.
	\end{equation}
	
	The helix is assumed to translate with an instantaneous velocity $U\zhat$ and to rotate with an instantaneous angular velocity $\omega \zhat$ in a Newtonian fluid of dynamic viscosity $\eta$, so that a point $t$ on the centreline has a local {instantaneous velocity} given  by	$\mathbf w(t) = (-a\omega\sin t, a\omega\cos t, U)$.  {This could be, for example, an artificial bacterial flagellum driven in rotation by an external field,  or the  free-swimming of a bacterial flagellar filament, or other related physical setups.}
	The helix is coupled to a viscous load taken for simplicity as a sphere of radius $R$. The Reynolds number is assumed to be much smaller than unity i.e.~the flows are Stokesian, as relevant for all applications cited above~\cite{lauga2020fluid}.

	In order to determine the hydrodynamic force density $\mathbf f(t)$ exerted by the filament on the fluid along the centreline of the helix, we use the slender-body theory (SBT) derived originally by Lighthill~\cite{lighthillsiamreview} taking the form of a non-local integral relation
	\begin{equation}\label{eq:SBT}
		\mathbf w(t_0) = \frac{\mathbf f_\perp(t_0)}{4\pi\eta} + \int_{r'>q_0}\frac{r'^2\mathbf f(t) + (\mathbf f(t) \cdot \mathbf r')\mathbf r'}{8\pi\eta r'^3}~\mathrm dt \sqrt{a^2 + b^2},
	\end{equation}
	where $q_0 = \frac{1}{2}r_0\sqrt{e}$, and $\mathbf r' = \mathbf r(t_0) - \mathbf r(t)$ is the position of the point at parameter $t_0$ on the helix relative to the point at $t$, with $r' = |\mathbf r'|$. If $\mathbf{\hat t}$ is the local unit vector tangent to the helical centreline, we denote the locally tangential and normal components of the force distribution $\mathbf f$ as $\mathbf f_\parallel := (\mathbf f \cdot \mathbf{\hat t}) \mathbf{\hat t}$ and $\mathbf f_\perp := \mathbf f - \mathbf f_\parallel$, respectively.

	\section{General SBT solution for force distribution}\label{sec:generalsolution}
	
	{Following the initial approach of Lighthill, we now take advantage of the helical symmetry to decompose the force distribution into components parallel and normal to the helical axis and simplify the SBT integral equation into a linear relation between the kinematics of these two components of force distribution. This is a general relation which lays the groundwork for our solutions in later sections for the more specific physical setups of force-free  swimming under an external torque in \S\ref{sec:swimmingproblems}\ref{sec:forcefree}   and force- and torque-free swimming in \S\ref{sec:swimmingproblems}\ref{sec:torquefree}. We first present a derivation of this relation, and then validate it numerically in the case of a helix rotating without translation.}

	\subsection{Semianalytical SBT solution}
	
	Using the helical symmetry of the problem in the $n \to \infty$ limit, we follow Ref.~\cite{Lighthill1996HelicalDO} and seek an ansatz  for $\mathbf f(t)$ of the form
	\begin{equation}\label{eq:ansatz}
		\mathbf f(t) = (h \sin t, -h \cos t, f).
	\end{equation}
	This approximation is exact for an infinite helix, and is expected to be valid in the bulk of the helix when $n$ is finite.
	By helical symmetry, it is sufficient to consider the SBT Equation~\eqref{eq:SBT} at $t_0 = 0$; this yields a linear system of equations which we can then solve for $h$ and $f$. We now calculate explicitly the various terms in the RHS of Equation~\eqref{eq:SBT}.
	
	The local tangent to the helical centreline is given by $\mathbf{\hat t}(t) = (-\cos \beta \sin t, \cos \beta \cos t, \sin \beta)$, where $\tan \beta := b/a$, yielding for the first term
	\begin{eqnarray}\label{eq:f0}
		\mathbf f_\perp(0) &= (0, -h\sin^2\beta - f \sin\beta\cos\beta, h \cos\beta\sin\beta + f \cos^2\beta).
	\end{eqnarray}
	
	We now evaluate the non-local contribution, i.e.~the integral term. Since $t_0 = 0$, $\mathbf r'(t) = (a - a \cos t, -a\sin t, -b t)$ and we have $r'(t)^2 = b^2t^2 + 2a^2(1-\cos t)$ and $\mathbf f(t)\cdot \mathbf r' = ah\sin t - bft$. Define $\delta$ to be the positive root of $r'(t) = q$ so that the integral is over $t \in [-n\pi,-\delta]\cup[\delta,n\pi]$.   
	
	The first term in the integral (without the prefactors) is, {integrating by parts,}
	\begin{subequations}
		\begin{align}\nonumber
			\int_{r'>q_0} \frac{\mathbf f(t)}{r'}\,\mathrm dt 
			&= \left[\frac{1}{r'}\begin{pmatrix}
				-h \cos t\\ -h \sin t\\ ft
			\end{pmatrix}\right]_{\delta_0}^{n\pi} + \left[\frac{1}{r'}\begin{pmatrix}
				-h \cos t\\ -h \sin t\\ ft
			\end{pmatrix}\right]_{-n\pi}^{-{\delta_0}} \\
			&+ \int_{r'>q_0}\frac{1}{r'^3}\begin{pmatrix}
				-b^2h t\cos t - a^2h \sin t \cos t \\ -b^2ht\sin t - a^2 h \sin^2 t \\ b^2ft^2 + a^2ft\sin t
			\end{pmatrix}\,\mathrm dt\\
			&= \frac{2}{b}\begin{pmatrix}
				0 \\ { -\frac{h\sin(n\pi)}{[n^2\pi^2 + 2\epsilon^2(1 - \cos n\pi)]^{1/2}}} + \frac{hb\sin \delta_0}{q_0} - h A_1 - \epsilon^2h A_2 \\ {\frac{fn\pi}{[n^2\pi^2 + 2\epsilon^2(1 - \cos n\pi)]^{1/2}}} - \frac{fb\delta_0}{q_0} + fA_3 + \epsilon^2f A_1
			\end{pmatrix},\label{eq:int1}
		\end{align}
	\end{subequations}
	{where the integrals $A_1, A_2$ and $A_3$ are defined as}
	{\begin{equation}
			(A_1, A_2, A_3) = \int_{\delta_0}^{\pi n} \frac{1}{(t^2 + 2\epsilon^2(1-\cos t))^{3/2}}(t\sin t, \sin^2 t, t^2 )~\mathrm d t.
	\end{equation}}
	
	The second term of the integral in Equation~\eqref{eq:SBT}     is {then}
	\begin{align}\label{eq:int2}
		\int_{r'>q_0}\frac{(\mathbf f(t)\cdot\mathbf r')\mathbf r'}{r'^3}\,\mathrm dt = \frac{2}{b}\begin{pmatrix}
			0 \\ -\epsilon^2hA_2 + \epsilon fA_1 \\ -\epsilon hA_1 + fA_3
		\end{pmatrix}.
	\end{align}
	
	{Using Equations~\eqref{eq:f0}, \eqref{eq:int1}, and \eqref{eq:int2}}, the SBT Equation~\eqref{eq:SBT} at $t_0 = 0$ is written as
	\begin{align}\label{eq:sbtmatrix}
		4\pi\eta\begin{pmatrix}
			a\omega \\
			U
		\end{pmatrix} = 
		\begin{pmatrix}
			-\Lambda_+&& \epsilon\Lambda_- \\
			-\epsilon\Lambda_- && \chi
		\end{pmatrix}
		\begin{pmatrix}
			h\\
			f
		\end{pmatrix},
	\end{align}
	where
	\begin{align}
		\lp &= -\sqrt{1 + \epsilon^2}\left({-\frac{\sin n\pi}{[n^2\pi^2 + 2\epsilon^2(1 - \cos n\pi)]^{1/2}}} + \frac{b}{q_0}\sin\delta_0 - A_1 - 2\epsilon^2A_2\right) + \frac{1}{1+\epsilon^2},\\
		\lm &= \sqrt{1 + \epsilon^2}A_1 - \frac{1}{1+\epsilon^2},\\
		\chi &= \sqrt{1 + \epsilon^2}\left({\frac{n\pi}{[n^2\pi^2 + 2\epsilon^2(1 - \cos n\pi)]^{1/2}}} - \frac{b}{q_0}\delta_0 +2A_3+\epsilon^2A_1\right)+\frac{\epsilon^2}{1+\epsilon^2}.
	\end{align}

	We have defined the entries of this matrix, as shall be shown later, such that $\Lambda_\pm$ and $\chi$ are positive $\mathcal O(\ln \eninv)$ quantities.

	%
	%
	%
	\subsection{Numerical validation}
	In this section we validate our semi-analytical solution against a numerical solution of SBT in the case of a helix rotating without translating. An external torque on the helix induces a rotation with a fixed angular velocity $\omega \zhat$ and an external force prevents translation.  Mathematically, we thus impose the physical conditions of prescribed $\omega$ with $U = 0$, in Equation~\eqref{eq:sbtmatrix}. Inverting the matrix yields a solution for $h$ and $f$, 
	\begin{equation}\label{eq:hf0}
		(h, f) = -\frac{4\pi\eta a \omega}{\lp\chi - \epsilon^2\lm^2} (\chi, \epsilon \lm)
	\end{equation}
	and the force distribution $\mathbf f(t)$ is as given in Equation~\eqref{eq:ansatz}.

	%
	

	A numerical solution for $\mathbf f(t)$ of the SBT Equation~\eqref{eq:SBT} is obtained by discretising the integral using the trapezoidal rule and solving the resulting linear system for the discretised force distribution. We avoid resolving to the small length scale $q_0$ by first analytically evaluating the contribution in $q_0 < r' < q$ to the SBT integrals using an asymptotic expansion, and then discretising the resulting equation. We require that the length scale $q$ is intermediate between the filament radius and the centreline curvature, i.e.~$r_0 \ll q \ll (a^2 + b^2)/a$; we henceforth take $q$ to be the geometric mean of this upper and lower bound.

	{In this paper, we  will  apply our results to the polymorphic forms of peritrichous bacteria such as \textit{E.~coli} and \textit{Salmonella}. Polymorphic forms are the distinct waveforms that these flagella may adopt, determined by the molecular structure of their protein subunits called flagellin \cite{namba1997}. Twelve polymorphic forms are thought to exist, including two straight forms \cite{calldine1978}. In Table~\ref{table}, we list the geometrical parameters corresponding to the ten helical forms. For each polymorphic form, the (dimensional) helical diameter $2a$ and helical pitch $2\pi b$ are averages over multiple values from the literature \cite{fujii2008,calldine1978,hasegawa1998,kamiya1976,darntonberg2007,iino1962,iino1966,hotani1982,iino1974,macnab1977,turner2000,asakura1970,matsuura1978}, experimentally measured for \textit{E.~coli} or \textit{Salmonella} or theorised, and collated in Tables S1, S2 and S4 of Ref.~\cite{lauga48}. The dimensionless parameters $\epsilon = a/b$, 	$\epsilon_0 = r_0/b$, and number of turns $n$ are calculated assuming a flagellar arclength $L = 10$ \textmu{m} and flagellar radius $r_0 = 10$~nm. In what follows, we will primarily focus on the most common form, known as the normal polymorph, highlighted in bold in Table~\ref{table}. }

	\begin{table}[!t]
		\begin{tabular}{llllll}
			\hline
			Polymorph&	$2a$ (\textmu{m})	&	$2\pi b$ (\textmu{m})	&	$\epsilon = a/b$	&	$\epsilon_0 = r_0/b$	&	$n$	\\
			\hline
			1 (Hyperextended)	&	0.08	&	1.58	&	0.16	&	0.040	&	6.3	\\
			\textbf{2 (Normal)}	&	\textbf{0.44}	&	\textbf{2.26}	&	\textbf{0.61}	&	\textbf{0.028}	&	\textbf{3.8}	\\
			3 (Coiled)	&	1.18	&	0.68	&	5.46	&	0.092	&	2.7	\\
			4 (Semi-Coiled)	&	0.55	&	1.20	&	1.46	&	0.053	&	4.7	\\
			5 (Curly I)	&	0.29	&	1.11	&	0.83	&	0.057	&	7.0	\\
			6 (Curly II)	&	0.15	&	0.92	&	0.51	&	0.068	&	9.7	\\
			7	&	0.05	&	0.70	&	0.22	&	0.089	&	13.9	\\
			8	&	0.03	&	0.60	&	0.13	&	0.105	&	16.5	\\
			9	&	0.01	&	0.52	&	0.06	&	0.121	&	19.2	\\
			10	&	0.16	&	1.01	&	0.50	&	0.062	&	8.9	\\
			\hline
		\end{tabular}
		\vspace*{-4pt}
		\caption{	\label{table}
			Geometrical parameters for different polymorphic forms of the flagellar filaments of peritrichous bacteria. Reported figures are	averages over multiple values from the literature~\cite{fujii2008,calldine1978,hasegawa1998,kamiya1976,darntonberg2007,iino1962,iino1966,hotani1982,iino1974,macnab1977,turner2000,asakura1970,matsuura1978}, as tabulated in Ref.~\cite{lauga48}. The normal polymorph, used throughout this article, is highlighted in bold.}
	\end{table}

\begin{figure}[t]\centering
		\includegraphics[width=\textwidth]{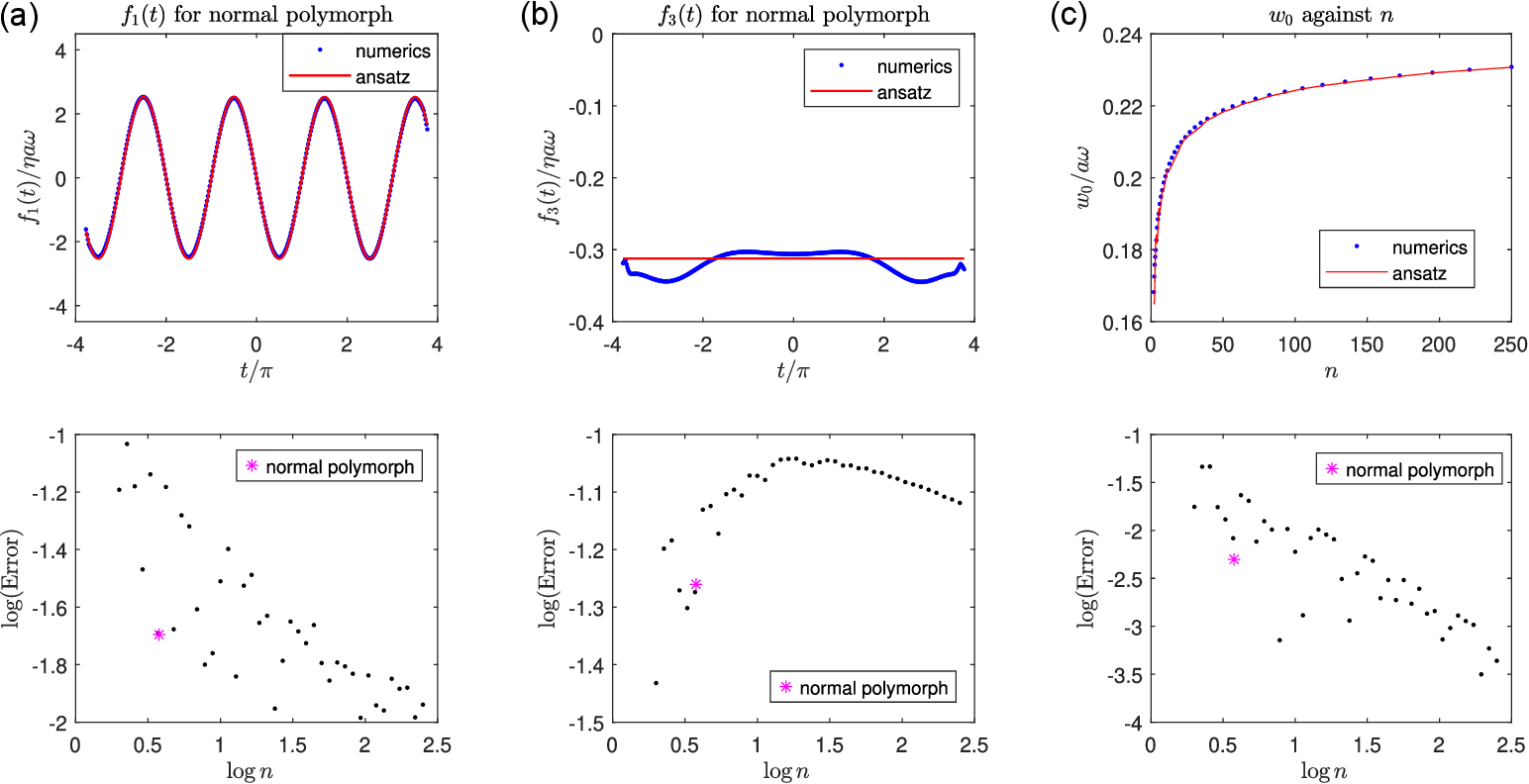}
		\caption{{{\bf Numerical validation of helical   ansatz}. A helical filament is rotated with an angular velocity $\omega \zhat$ while an external force prevents net translation. Top row: (a,b): The $x$-component (a) and $z$-component (b) of the force distribution $\mathbf f(t)$ along the centreline, and the $z$-component (c) of the flow field at the origin, $w_0$, induced by the rotating helix  as calculated using the numerical solution of SBT (blue) and   semi-analytically using the helically symmetric ansatz (red) are plotted in the top rows for the geometrical parameters $\epsilon = 0.61, \epsilon_0 = 0.028, n = 3.8$ corresponding to a normal polymorph.	 In the bottom row, the corresponding errors incurred by the theoretical ansatz, for  $\epsilon = 0.61, \epsilon_0 = 0.028$  plotted against the number of helical turns, $n$. In each panel the magenta asterisk denotes parameters corresponding to the normal polymorph.}}\label{fig:ansatz}
	\end{figure}

	We now present the results of the numerical validation in  {the top row of} Fig.~\ref{fig:ansatz}a-b where we plot the  theoretical results (red) for the $x$- and $z$- components of $\mathbf f(t)$  against numerical solutions (blue) for a helix with geometrical parameters $\epsilon = 0.61, \epsilon_0 = 0.028, n = 3.8$ corresponding to a normal polymorph (i.e.~with $2a = 0.44$~{\textmu}m, $2\pi b = 2.26$~{\textmu}m, $r_0 = 10$~nm, flagellum length $L = 10$~{\textmu}m). In the rest of the paper, we will use in our results this set of geometrical parameters unless stated otherwise.  {We obtain excellent agreement between the theoretical and numerical results for the $x$-component (Fig.~\ref{fig:ansatz}a). Discrepancies between theory and numerics are larger for the $z$-component (Fig.~\ref{fig:ansatz}b), but the theoretical approximation still performs well, in particular showing very good agreement with numerical results in the bulk of the helix away from the ends.}


	We quantify how close our theoretical solution is to the numerical solution by defining an RMS error for the $i^{\text{th}}$ component of $\mathbf f$ as 
	\begin{equation}
		\Delta_{i} = \left\{\frac{\int_{-n\pi}^{n\pi}[f_i^{\text{num}}(t) - f_i^{\text{theo}}(t)]^2~\mathrm dt}{\int_{-n\pi}^{n\pi}f_i^{\text{num}}(t)^2~\mathrm dt}\right\}^{\frac{1}{2}},
	\end{equation} 
	and plot these errors against $n$ in  {the bottom row of} Fig.~\ref{fig:ansatz}a-b, for the $x$- and $z$-components respectively. These errors remain small for $n$  as low as $n = 3.8$, relevant to the normal polymorph, indicated by magenta asterisks on those graphs {($\Delta_1 = 2.1\%, \Delta_3 =  5.5\%$)}.
	
	Finally, we calculate the $z$-component $w_0$ of the flow field $\mathbf u_0$ at the origin induced by the rotating helix, a measure of the flow `pumped' in the helix. This is obtained by integrating the Stokelets along the helix as follows:
	\begin{equation}\label{eq:u0}
		\mathbf u_0 = \int_{-\pi n}^{\pi n}\frac{r'^2\mathbf f(t) + (\mathbf f(t') \cdot \mathbf r')\mathbf r'}{8\pi\eta r'^3}~\mathrm dt \sqrt{a^2 + b^2},
	\end{equation}
	where now $\mathbf r' = -(a\cos t, a \sin t, bt)$ and $r' = |\mathbf r'|$. We numerically integrate Equation~\eqref{eq:u0}, using either the numerical SBT solution or the semi-analytical theoretical solution  for $\mathbf f(t)$, and plot $w_0$ against $n$ in Fig.~\ref{fig:ansatz}c (top). The error{, defined simply as $\Delta_w = |w_0^{\text{num}} - w_0^{\text{theo}}|/|w_0^{\text{num}}|$,} is again small for the value of $n$ relevant to the normal polymorph (magenta asterisk{, $\Delta_w = 0.5\%$}) and decays as $n \to \infty$ as expected,  {as shown in Fig.~\ref{fig:ansatz}c (bottom)}.
	
	{The dominant polymorphic forms of peritrichous bacteria are the normal, semi-coiled, and curly I polymorphs. We can repeat this analysis with the geometrical parameters of the remaining two polymorphs, and calculate the errors $\Delta_1 = 6.9\%, \Delta_3 = 14.3\%, \Delta_w = 3.8\%$ for the semi-coiled polymorph, and $\Delta_1 = 4.3\%, \Delta_3 = 10.3\%, \Delta_w = 0.7\%$ for the curly I polymorph. Since the value of $\epsilon$ is larger for these polymorphic forms than the normal form, the corresponding errors are also greater, but   the theoretical solution continues to largely agree with numerics for different geometrical parameters. It  is in fact surprising  how well the theoretical solution performs for the semi-coiled polymorph, considering that $\epsilon$ is greater than one in that case.}

	\section{SBT solution to swimming problems}

	\label{sec:swimmingproblems}
	We now investigate the rotation-driven swimming of a helical filament attached to a spherical body by imposing further physical conditions on the general solution  just derived, Equation~\eqref{eq:sbtmatrix}. We first solve in \S4\ref{sec:forcefree} the problem of force-free swimming of a helix rotated by an external torque, relevant to artificial bacterial flagella{, where this torque is applied via an external magnetic field}~\cite{zhang2009, zhang2010} (Fig.~\ref{fig:swimming}a). We then solve in \S4\ref{sec:torquefree} for the dynamics of a free-swimming helical swimmer, relevant to the motility of flagellated bacteria~\cite{wadhwa2022, berg2004, lauga2016}  (Fig.~\ref{fig:swimming}b).

	\subsection{Force-free swimming under external torque}\label{sec:forcefree}

	\subsubsection{Solution}
\begin{figure}[t]\centering
		\includegraphics[width=\textwidth]{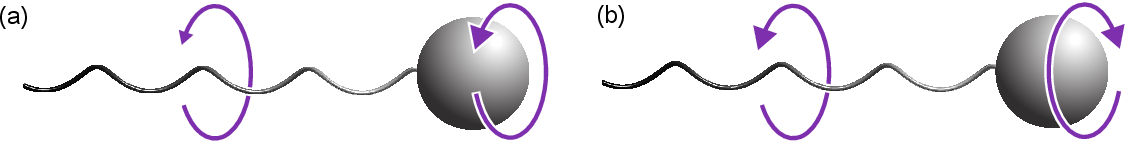}
		\caption{{\bf Swimming problem setups}. (a) A helical filament attached to a spherical load subject to an external torque;   the filament and load have no relative motion and they co-rotate by translating. (b) A force- and torque-free helical swimmer in which specialised rotary  motors at the filament-load junction cause the filament and body to undergo relative rotation; the load and filament always counter-rotate so that the swimmer remains overall torque-free.}\label{fig:swimming}
	\end{figure}

	In this first case, an external torque is applied onto the helix so that it rotates with   angular velocity $\omega \zhat$. The helix is attached to a spherical head of radius $R$ and there is no relative rotation between the helix and the load (Fig.~\ref{fig:swimming}a). The helix is thus propelled at some velocity $\mathbf U$ such that the whole swimmer experiences zero net force at each time. 
	
	Mathematically we are therefore prescribing the angular velocity $\omega$ and imposing the force-free condition
	\begin{equation}\label{eq:forcefund}
		\int_{-n\pi}^{n\pi} \mathbf f(s)\,\mathrm ds + 6\pi\eta R \mathbf U = \boldsymbol 0,
	\end{equation}
	where the two terms represent the forces exerted  on the fluid by the helix  and the cell body respectively. With our helically symmetric approximation Equation~\eqref{eq:ansatz} for $\mathbf f$, this simplifies, when $n$ is an integer, to $\mathbf U = U \zhat$ (i.e.~straight swimming along the helical axis) with
	\begin{equation}\label{eq:forcebalance}
		U = -\frac{nf\sqrt{a^2 + b^2}}{3\eta R}.
	\end{equation} 
	When $n$ is not an integer, our ansatz cannot satisfy the $y$ component of Equation~\eqref{eq:forcefund} i.e.~a straight-swimming helix exerts a non-zero  {time-varying force onto the fluid (in the lab frame)}, but we will justify later that this force is small i.e.~a straight-swimming helix is to a good approximation force-free, and this equation remains valid to a good approximation for non-integer values of $n$.
	
	We may next use the force balance condition, Equation~\eqref{eq:forcebalance}, to eliminate $U$ in Equation~\eqref{eq:sbtmatrix}, leading to
	\begin{align}\label{eq:Uelim}
		\begin{pmatrix}
			4\pi\eta a\omega \\
			0
		\end{pmatrix} = 
		\begin{pmatrix}
			-\lp && \epsilon \lm \\
			-\epsilon \lm && \chi + \frac{4\pi n \sqrt{a^2 + b^2}}{3R}
		\end{pmatrix}
		\begin{pmatrix}
			h\\
			f
		\end{pmatrix}.
	\end{align}
	Inverting this for $f$ and using Equation~\eqref{eq:forcebalance}, the swimming speed is finally obtained explicitly  as
	\begin{equation}\label{eq:Uforcefree}
		U = \frac{a\omega \epsilon \Lambda_-}{\displaystyle \Lambda_+\left(1 + \frac{3\chi R}{4
				\pi n \sqrt{a^2 + b^2}}\right) - \epsilon^2\frac{3 \lm^2 R}{4\pi n\sqrt{a^2 + b^2}}}.
	\end{equation} 
	
	%
	Note that in the limit $R \to \infty$, we obtain $U \to 0$ and we recover the solution, Equation~\eqref{eq:hf0}, for a torqued helix prevented from translating.

	\subsubsection{Justifying the straight-swimming approximation for $n \notin \mathbb N$}\label{sec:justifyforce}
	Here we consider the limit in which  a straight-swimming helix may be force-free to a good approximation  even when $n $ is not an integer.
	
	A helix rotating about the $z$ axis exerts onto the fluid a $y$-force equal to $-2bh \sin \pi n$, which is of order $bh$ in the worst case scenario. To balance this the swimmer translates in the $y$ direction at a velocity $V$. Using resistive-force theory we may make an order-of-magnitude estimate of the force density along the helix required to sustain this translation as order $\eta V/\ln \eninv$, corresponding to a net force onto the fluid from the helix of order $\eta VL/\ln \eninv$. The force onto the fluid from the $y$-translation of the cell body is of order $\eta RV$. Force balance on the swimmer therefore reads $bh = \mathcal O(\eta VL/\log \eninv) + \mathcal O(\eta RV)$,	so
	\begin{equation}
		V \sim \frac{bh}{\eta\left(R + L/\log \eninv\right)} < \frac{bh \log \eninv}{\eta L }.
	\end{equation}
	This gives us an order of magnitude estimate of the speed at which the helix translates in order to balance the $y$ component of force. This corresponds to a force density of $\sim \eta V/\log \eninv \sim hb/L$; this is the size of the correction needed to $h$ in the straight-swimming case in order to account for the $y$ translation in the truly force-free motion.  {The relative correction needed to $h$ is therefore order $b/L$; for a bacterial flagellum in its normal polymorphic form ($b = 0.36$ \textmu{m}, $L = 10$ \textmu{m}), this quantity has a numerical value of 4\%. The straight-swimming assumption is therefore reasonable for the dimensions of bacterial flagella.} {Experimental \cite{peyer2010} and theoretical\cite{man2013} studies which account more realistically for the external magnetic torque reveal more interesting dynamics, such as wobbling about the swimming direction at magnetic field frequencies which are too low or too high, but we have shown here that the rotation $\omega \zhat$ of a helical filament about the helix axis, to a good approximation, results in straight-swimming.}

	\subsection{Force- and torque-free swimming}\label{sec:torquefree}
	\subsubsection{Torque balance}
	
	We now consider the problem, motivated by the locomotion  of flagellated bacteria, of the force- and torque-free motion of a helical filament   attached to a sphere of radius $R$. The balance of torques in this case is enabled by the counter-rotation of the body relative to the filament (Fig.~\ref{fig:swimming}b).

	The torque  $\boldsymbol \tau_{\mathrm{helix}}$ exerted by a helix with a force distribution $\mathbf f(t) = (h \sin t, -h \cos t, f)$ on the fluid is
	\begin{align}\nonumber
		\boldsymbol \tau_{\mathrm{helix}} &= \int_{-n\pi}^{n\pi}
		\begin{pmatrix}
			a \cos t \\ a \sin t \\ bt
		\end{pmatrix}
		\times
		\begin{pmatrix}
			h \sin t \\ - h \cos t \\ f
		\end{pmatrix}
		\sqrt{a^2 + b^2}\,\mathrm d t \\ 
		&=\sqrt{a^2 + b^2}	\begin{pmatrix}
			0 \\ bh(2 \sin \pi n - 2\pi n \cos \pi n) - 2af \sin \pi n \\ -2\pi n ah
		\end{pmatrix},
	\end{align}
	while the torque exerted on the fluid by a sphere of radius $R$ which rotates with angular velocity $\boldsymbol \Omega$ is 
	$\boldsymbol \tau_{\mathrm{body}} = 8\pi\eta R^3 \boldsymbol\Omega$. The overall torque balance requires 
	\begin{equation}\label{eq:torquefund}
		\boldsymbol\tau_{\mathrm{helix}} + \boldsymbol\tau_{\mathrm{body}} = \boldsymbol 0.
	\end{equation}
	
	The $z$ component of $\boldsymbol \tau_{\mathrm{helix}}$ can be balanced by a counter-rotation of the cell body about the $z$ axis. In order to balance the $y$ component, a rotation of the swimmer about the $y$ axis is required, which breaks the helical symmetry and makes SBT analytically intractable.  However, we proceed with the assumption,  justified in \S4(b)\ref{sec:justifytorque}, that any correction to the straight-swimming case is small i.e.~a force- and torque-free helical swimmer has, to a good approximation, its translational and angular velocities in the $\zhat$ direction.
	
	\subsubsection{Solution for prescribed rotation rate}
	We first consider the case in which the helical filament rotates with a prescribed angular velocity $\omega'\zhat$ relative to the load (i.e.~the cell body). The load counter-rotates with an angular velocity $\Omega$ so that the helix rotates with a net angular velocity $\omega \zhat = (\omega' + \Omega)\zhat$ relative to the fluid at infinity. As a result of this rotation, swimming occurs instantaneously with velocity $U \zhat$.

	Torque balance, Equation~\eqref{eq:torquefund}, in the $z$-direction yields
	\begin{equation}\label{eq:torquebalance}
		\Omega = \frac{a\sqrt{a^2 + b^2}nh}{4\eta R^3},
	\end{equation}
	while Equation~\eqref{eq:forcebalance} for the force balance remains valid. Using these two equations, we may eliminate $U$ and $\Omega$ in Equation~\eqref{eq:sbtmatrix}, yielding
	\begin{align}\label{eq:forcetorquefreematrix}
		\begin{pmatrix}
			4\pi\eta a\omega' \\
			0
		\end{pmatrix} = 
		\begin{pmatrix}
			-\lp - \frac{\pi a^2 \sqrt{a^2 + b^2}n}{R^3} && \epsilon \lm \\
			-\epsilon \lm && \chi + \frac{4\pi \sqrt{a^2 + b^2}n}{3R}
		\end{pmatrix}
		\begin{pmatrix}
			h\\
			f
		\end{pmatrix}.
	\end{align}
	Similarly to before, we may solve this for $f$ and use the force-free condition, Equation~\eqref{eq:forcebalance}, to obtain  the swimming speed $U$ as
	\begin{equation}\label{eq:Utorquefree}
		U = \frac{a\omega \epsilon \Lambda_-}{\displaystyle\left(\Lambda_+ + \frac{\pi n a^2 \sqrt{a^2 + b^2}}{R^3}\right)\left(1 + \frac{3\chi R}{4
				\pi n \sqrt{a^2 + b^2}}\right)- \epsilon^2\frac{3 \lm^2 R}{4\pi n\sqrt{a^2 + b^2}}}.
	\end{equation}
	The torque-free condition, Equation~\eqref{eq:torquebalance}, together with the solution of Equation~\eqref{eq:forcetorquefreematrix} yields the angular velocity of the cell body's counter-rotation as
	\begin{equation}
		\Omega = -\frac{\omega'}{\displaystyle1 + \frac{\lp R^3}{\pi a^2 \sqrt{a^2 + b^2} n}}.
	\end{equation} 
	Note that as $R \to 0$, $\omega = \Omega + \omega'\to 0$: therefore a helix with no load cannot rotate and remain torque-free, and there is no propulsion without this rotational driving. In the other limit, $R \to \infty$,  there is also no propulsion simply because the viscous drag is too large.

	\subsubsection{Solution for prescribed motor torque}
	{Biologically, the rotation of a bacterial flagellar filament is powered by the torque exerted onto it by a specialised rotary motor at its base~\cite{berg1993}. A biologically relevant question is therefore to  investigate how the swimming speed varies with the load for a prescribed motor torque}. In what follows, instead of prescribing $\omega'$, we therefore prescribe the value of the torque $G \zhat$  on the helix. In this case torque balance on the helix yields
	$	2\pi n a \sqrt{a^2 + b^2}  h + G = 0$.
	, which determines $h$ as 
	\begin{equation}\label{eq:hmotortorque}
		h = -\frac{G}{2a\sqrt{a^2 + b^2}\pi n}.
	\end{equation}
	Eliminating $U$ using force balance as before yields  Equation~\eqref{eq:Uelim}, from which we can derive the value of $f$ as
	\begin{equation}\label{eq:fmotortorque}
		f = \frac{\epsilon\Lambda_- h}{\displaystyle \chi + \frac{4\pi \sqrt{a^2 + b^2} n}{3R}} = -\frac{\epsilon G \Lambda_-}{\displaystyle  2a\sqrt{a^2 + b^2}\pi n \left(\chi + \frac{4\pi\sqrt{a^2 + b^2} n}{3R}\right)},
	\end{equation}
	and   the rotation rate $\omega$ of the helix in the lab frame,
	\begin{equation}\label{h2}
		\omega = -\frac{\lp h}{4\pi\eta} = \frac {\Lambda_+G}{\displaystyle 8\pi^2 \eta na^2\sqrt{a^2 + b^2}}.
	\end{equation}
	The swimming speed may finally be calculated as before from the force balance condition as
	\begin{equation}
		U = \frac{\epsilon G\Lambda_-}{\displaystyle 8\pi^2a\sqrt{a^2 + b^2}\eta n \left(1 + \frac{3\chi R}{4\pi \sqrt{a^2 + b^2} n}\right)}.
	\end{equation}
	Note that in this case we have a finite swimming speed as $R \to 0$ since  a non-vanishing torque is applied onto the helix.

	\subsubsection{Justifying the straight-swimming approximation}\label{sec:justifytorque}
	We now justify our assumption that a straight-swimming motion is a good approximation of a force- and torque-free motion. We have already shown in \S4(a)\ref{sec:justifyforce} that a straight-swimming helix is force-free up to $\mathcal O(b/L)$ corrections; here we show that it is also torque-free up to $\mathcal O(b/L)$ corrections.

	A swimmer rotating about the $z$ axis will in the worst case scenario exert onto the fluid a torque in the $y$ direction of order $2\pi nb^2h$. To balance this the cell body rotates about the $y$ axis at an angular speed $\Omega_2$. 
	This induces a velocity of order $\Omega_2 L$ at a typical point on the helix. Using resistive-force theory we may make an order-of-magnitude estimate of the force density along the helix required to sustain this motion as $\sim \eta \Omega_2 L/\log \eninv$, corresponding to a torque onto the fluid from the helix of order $\eta \Omega_2 L^3 /\log \eninv $. 	The torque onto the fluid from the $y$-rotation of the cell body is order $\eta R^3 \Omega_2$. Torque balance on the swimmer then reads $2\pi nb^2 h  = \mathcal O({\eta \Omega_2 L^3}/{\log \eninv}) +\mathcal O (\eta R^3 \Omega_2)$, so
	\begin{equation}
		L\Omega_2 \sim \frac{bh}{\eta\left({L}/{\log\eninv} + {R^3}/{L^2}\right)} < \frac{bh \log \eninv}{\eta L }.
	\end{equation}
	We now have an expression for $L\Omega_2$, which is the typical speed of a point on the helix associated with the motion needed to balance $y$ component of torque. This corresponds to a force density of order $L\Omega_2\eta/\log\eninv \sim hb/L$; this is the size of the correction needed to $h$ in the straight-swimming case in order to account for the $y$-rotation in the truly torque-free case.   {We again see that a relative correction of $b/L$ is needed, which is around 4\% for the normal polymorphic form of a peritrichous bacterial flagellum.} While a real bacterium does exhibit slight wobbling about its swimming direction and therefore an overall helical swimming trajectory, a subject of more detailed hydrodynamic studies \cite{darnton2007,hyon2012}, the straight-swimming assumption is reasonable and rigorously justified for the purposes of deriving an approximate solution in the style of Lighthill.

	\section{Resistive-force theory (RFT) solutions}\label{sec:rft}
	
	\subsection{RFT solution for the force distribution}
	The problems in the previous section may instead be solved using using resistive-force theory, an approximation to slender-body theory accurate to order $\log \epsilon$. In this framework the force $\mathbf f$ the filament exerts onto the fluid obeys an anisotropic proportionality relationship with the local filament velocity. This may be written at a point $t_0$ on the helix as 
	\begin{equation}\label{eq:rft}
		\mathbf w(t_0) = \alpha_\perp \mathbf f_\perp(t_0) + \alpha_\parallel \mathbf f_\parallel(t_0),
	\end{equation}
	where the subscripts denote components in the directions locally normal/parallel to the filament, and the proportionality constants $\alpha_\perp$ and $\alpha_\parallel (>\alpha_\perp)$ are the mobility coefficients. Under RFT, the ansatz for $\mathbf f$ in Equation~\eqref{eq:ansatz} is exact for any value of $n$.  For a helix    with translational and angular velocities $U \zhat$ and $\omega \zhat$, respectively,	this then yields the linear system
	\begin{align}\label{eq:rftmatrix}
		\begin{pmatrix}
			a\omega \\
			U
		\end{pmatrix}
		=\frac{1}{1 + \epsilon^2}
		\begin{pmatrix}
			-\alpha_\perp - \epsilon^2 \alpha_\parallel  & (-\alpha_\perp + \alpha_\parallel)\epsilon \\(\alpha_\perp - \alpha_\parallel)\epsilon & \alpha_\parallel+\epsilon^2 \alpha_\perp
		\end{pmatrix}
		\begin{pmatrix}
			h\\f
		\end{pmatrix},
	\end{align}
	which is analogous to Equation~\eqref{eq:sbtmatrix} obtained using SBT. Note, however, that unlike Equation~\eqref{eq:sbtmatrix}, the result in Equation~\eqref{eq:rftmatrix} is an exact equality under RFT. This may be coupled with other physical conditions to solve for the force distribution, similarly to how we have used  Equation~\eqref{eq:sbtmatrix} in the previous sections.
	
	\subsection{Issues with Lighthill's optimal RFT}

	\subsubsection{Lighthill's coefficients}
	In the simple result in Equation~\eqref{eq:rftmatrix},  
	The question arises of what values of the mobility coefficients ($\alpha_\perp, \alpha_\parallel$) should be used. Lighthill derived, \textit{for the case of an infinite helix rotating freely without a load}, the `optimal' values of these coefficients,
	\begin{equation}
		\alpha_\perp^{\mathrm{Lighthill}} = \frac{ {1}/{2} + \log\left({0.18\ell}/{r_0}\right)}{4\pi\eta}, 	\,\,\,	\alpha_\parallel^{\mathrm{Lighthill}} = \frac{\log\left( {0.18\ell}/{r_0}\right)}{2\pi\eta},
	\end{equation} 
	where $\ell = 2\pi\sqrt{a^2 + b^2}$ is the wavelength of the helix along the curved centreline. With these values, the prediction from RFT yields the same result  as that from SBT~\cite{lighthillsiamreview}.

	However, these Lighthill coefficients have since been used ubiquitously in the literature~\cite{peyer2013,martindale2016,wang2017,palusa2018}, regardless of whether the conditions under which these coefficients were originally derived hold, sometimes even criticised for inaccuracy despite usage outside their regime of validity \cite{rodenborn2013}.

	Many problems of interest in the microswimmer literature involve a viscous load attached to a helical filament (e.g.~a bacterial cell body, or a cargo for artificial swimmers). It is therefore mathematically inconsistent in these cases to use Lighthill's coefficients which were derived for a free helix. In this section, we  demonstrate   that using Lighthill's coefficients outside their regime of validity leads to significant physical/numerical inaccuracies. Specifically we consider  the flows created by a rotating helix, and  the swimming speeds of a helical microswimmer.

	\subsubsection{Flow field induced by a rotating helix}\label{sec:flowfieldinhelix}
	
	We first consider the flow field induced in the $z = 0$ plane by a helical filament which rotates with an angular velocity $\omega \zhat$ and does not translate. 	The flow velocity $\mathbf u(x,y,0) = (u(x,y),v(x,y),w(x,y))$ is obtained by integrating the Stokeslets:
	\begin{equation}\label{eq:uxy0}
		\mathbf u(x,y,0) = \int_{-\pi n}^{\pi n}\frac{r'^2\mathbf f(t) + (\mathbf f(t') \cdot \mathbf r')\mathbf r'}{8\pi\eta r'^3}~\mathrm dt \sqrt{a^2 + b^2},
	\end{equation}
	where $\mathbf r'(t) = (x - a\cos t, x - a\sin t, -bt)$ and $r' = |\mathbf r'|$. 
	We then compute $\mathbf u(x,y,0)$ by numerically integrating Equation~\eqref{eq:uxy0}, using either the RFT solution (specifically, we impose $U = 0$ in Equation~\eqref{eq:rftmatrix} to deduce the RFT value of $\mathbf f(t)$) or the numerical SBT solution for $\mathbf f(t)$.
	
\begin{figure}[t]\centering
		\includegraphics[width=\textwidth]{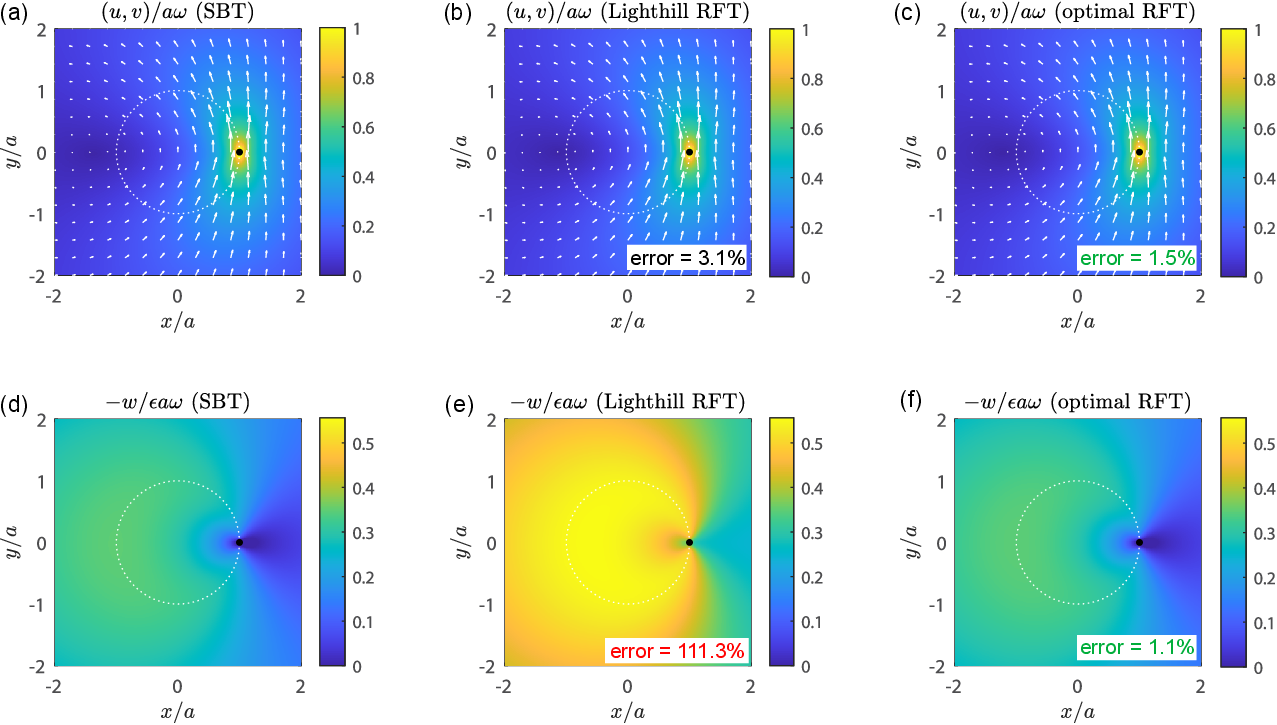}
		\caption{{\bf Flow field induced by a rotating helix.} $(x,y)$-component (top; a-c) and $z$-component (bottom; d-f) of the flow field, normalised as indicated in panel headings, in the $z = 0$ plane of a helical filament rotating at angular velocity $\omega \zhat$ and prevented from undergoing translation, as calculated using SBT (left; a,d), RFT with Lighthill's coefficients (middle; b,e), and RFT with optimal coefficients (right; c,f) respectively. Geometrical parameters used are $\epsilon = 0.61, \epsilon_0 = 0.028, n = 3.8$ and correspond to a normal polymorph. Insets in (b,c,e,f) indicated maximum errors relative to SBT solution, as defined in main text.}\label{fig:flowinhelix}
	\end{figure}

	The comparison between the two is shown in 
	Fig.~\ref{fig:flowinhelix}, where we plot the components $(u,v)$ (perpendicular to helix axis, top row) and $w$ (along the helix axis, bottom row) of the flow field $\mathbf u(x,y,0)$ in the $z = 0$ plane, as calculated using SBT (leftmost column) or RFT with Lighthill's coefficients (middle column).  {The rightmost column then shows results calculated using a new  `optimal' RFT, which we   derive and interpret   in \S\ref{sec:optimalcoeff}.}  {We see that Lighthill's RFT yields similar results to SBT for the components $(u,v)$ (Fig.~\ref{fig:flowinhelix}a-b) but severely overestimates the value of $w$ (Fig.~\ref{fig:flowinhelix}d-e).}
	
	We quantify the maximum errors incurred by using RFT as follows. 
	For $(u,v)$,
	we define an error $E_{1,2} = \max_{x,y}|\mathbf u_H^{\text{SBT}}(x,y,0) - \mathbf u_H^{\text{RFT}}(x,y,0)|/a\omega$, where $\mathbf u_H = u\xhat + v\yhat$. The error is normalised by the instantaneous speed $a\omega$ of the filament. The error in $w$ is defined as $E_3 = \max_{x,y}|w^{\text{SBT}}(x,y,0) - w^{\text{RFT}}(x,y,0)|/w_{\text{max}}$, with $w_\text{max} = \max_{x,y}|w^{\text{SBT}}(x,y,0)|$.  		Lighthill's RFT incurs an overall error of $E_{1,2} = $ 3.1\% for the $(x,y)$-component, but predicts for the $z$-component a flow that is twice as strong as the SBT flows, leading to an error of $E_3 = 111.3$\%. This has dramatic implications for the use of Lighthill's RFT coefficients to calculate, for instance, the flows `pumped' by an artificial bacterial flagellum~\cite{zhang2010}.
	
	The relatively low errors in $(u,v)$ may be rationalised by  {asymptotically expanding the integral in Equation~\eqref{eq:uxy0} using $\epsilon$ as a small parameter, keeping the dependence on $h$ and $f$ explicit, to obtain} 	\begin{subequations}
		\begin{align}\label{eq:asym1}
			u(x,y,0) =& \frac{h}{4\pi\eta}\frac{(1 - \tilde x)\tilde y}{(1 - \tilde x)^2 + \tilde y^2 }, \\ \label{eq:asym2}
			v(x,y,0) =& -\frac{h}{4\pi\eta}\left\{\log\epsilon^{-1}  -\frac{1}{2}\log[(1-\tilde x)^2 + \tilde y^2] -\frac{\tilde y^2}{(1 - \tilde x)^2 + \tilde y^2.} + \log 2 -\gamma  + \frac{\sin \pi n}{\pi n}-\frac{\cos\pi n}{\pi^2 n^2}\right\},\\ \nonumber
			w(x,y,0) =& \frac{f}{4\pi\eta}\left\{2\log \epsilon^{-1} + 2\log \pi n + 2\log 2 - 1 - \log[(1 - \tilde x)^2 + \tilde y^2] \right\} \\
			&-\frac{\epsilon h}{4\pi\eta}\left\{\left[\log \epsilon^{-1} + \log 2 - \gamma - \log\sqrt{(1 - \tilde x)^2 + \tilde y^2}\right] \tilde x + \frac{\tilde y^2}{(1 - \tilde x)^2 + \tilde y^2 }\right\}, 
		\end{align}
	\end{subequations}
	where $(\tilde x, \tilde y) = (x/a, y/a)$. 	 {Here we have expanded $(u,v)$ to order $\epsilon^0 h/\eta$ and $w$ (which scales as $\epsilon u$) to  order $\epsilon h/\eta$.  	We see that to this order, $(u,v)$ depends only on $h$, whereas since $f \sim \epsilon h$, the dominant contribution to $w$ depends on both $h$ and $f$.} The fundamental issue with Lighthill's coefficients appears therefore to be that they do not account correctly for the force balance in the $z$ direction, but this has little effect on the $(x,y)$ components of the force distribution, which is captured by $h$. The effect of this error is seen most prominently in $f$, and therefore in $w$.

	\subsubsection{Swimming speeds of a helical swimmer}\label{sec:swimmingspeeds}
	
	We now present the RFT solution of the problem, introduced in \S4\ref{sec:torquefree}, of a force- and torque-free swimmer whose helical filament rotates at a fixed angular velocity $\omega'\zhat$ relative to the cell body. Since we are now using RFT, we can start with Equation~\eqref{eq:rftmatrix} instead of the analogous equation (Equation~\ref{eq:sbtmatrix}) for SBT. The rest of the calculation follows as in \S4\ref{sec:torquefree}  to yield the free swimming speed
	\begin{equation}\label{eq:UtorquefreeRFT}
		{U^{RFT} = \frac{(1+\epsilon^2)(\alpha_\parallel - \alpha_\perp)}{\displaystyle \frac{6\pi\eta R}{L}\left[\displaystyle \left(\alpha_\perp + \epsilon^2 \alpha_\parallel + \frac{(1 + \epsilon^2)a^2L}{8\pi\eta R^3}\right)\left(\alpha_\parallel + \epsilon^2\alpha_\perp + \frac{(1 + \epsilon^2)L}{6\pi\eta R}\right) - \epsilon^2(\alpha_\perp - \alpha_\parallel)^2\right]}\epsilon a\omega,}
	\end{equation}
	{where $L = 2\pi n/\sqrt{a^2 + b^2}$ is the arclength of the helix.}

	\begin{figure}[t]\centering
		\includegraphics[width=0.8\textwidth]{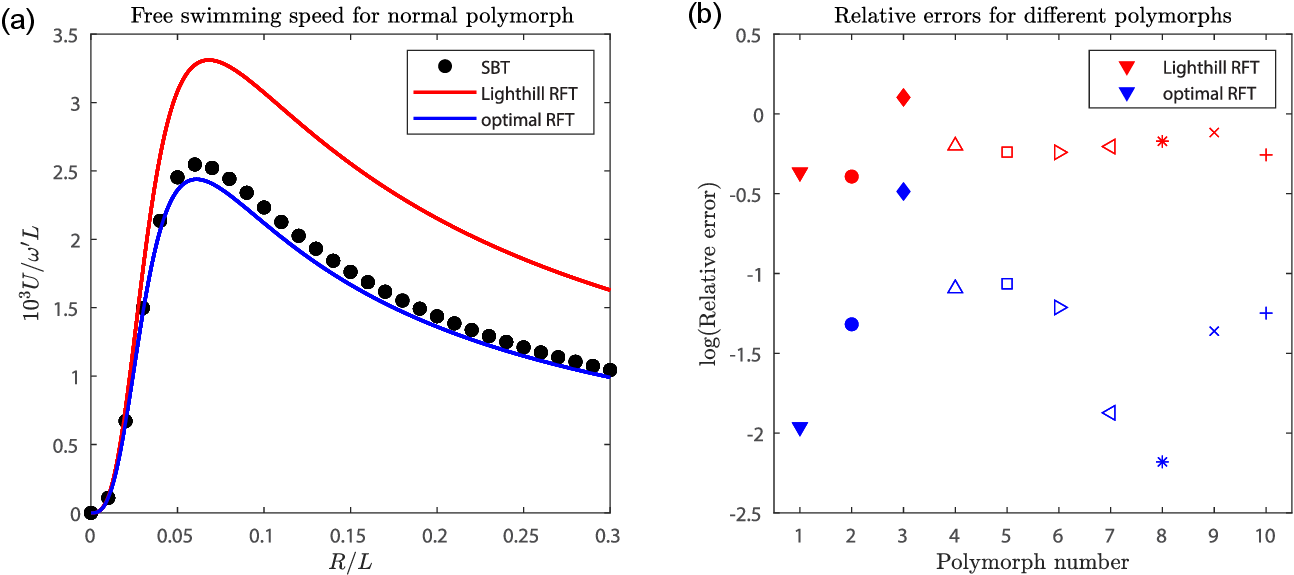}
		\caption{{\bf SBT vs RFT predictions for swimming speeds.}  (a) Swimming speeds (normalised) against load radius $R$ for  force- and torque-free swimming with a helical filament rotating with  fixed angular velocity $\omega \zhat$ relative to the cell body, as calculated numerically using SBT (black) and analytically using RFT with Lighthill's coefficients (red) and RFT with our optimal coefficients (blue). (b) Relative errors between RFT and  SBT incurred by using RFT with Lighthill's coefficients (red) and our
			optimal coefficients (blue) for the free swimming problem in
			(a), with the geometrical parameters of different bacterial polymorphs  {(see Table~\ref{table}).} {The same symbols as in Ref.~\cite{lauga48} are used  for the different polymorphs, with filled  markers indicating left-handed helices and empty markers for right-handed.}}\label{fig:swimmingspeeds}
	\end{figure}
	
	We may also numerically calculate the swimming speed as follows. We consider first a swimmer, whose filament rotates at an angular velocity $\omega' \zhat $ relative to the cell body, undergoing a rigid body motion $(\mathbf U, \Omega)$. We define $\mathbf H(\mathbf U, \Omega) = (\mathbf F, \mathbf G)$, the instantaneous net force and torque exerted onto the fluid by such a swimmer, which we can compute numerically using SBT and the hydrodynamic resistances of a spherical load (i.e.~cell body). The motion of the swimmer under zero net force and torque is obtained by solving $\mathbf H(\mathbf U, \Omega) = \boldsymbol 0$, and we extract from this a `forward' swimming speed given by the $z$-component of $\mathbf U$.

	
	These numerically obtained swimming speeds are plotted in Fig.~\ref{fig:swimmingspeeds}a, as black dots, for different values of the load radius $R$, together with the prediction from Lighthill's RFT in red (Equation~\eqref{eq:UtorquefreeRFT}).
	We see that Lighthill's RFT consistently overestimates swimming speeds,  {by up to nearly 50\% for larger loads}.

	\section{Optimal resistance coefficients}\label{sec:optimalcoeff}
	
	We have shown that Lighthill's RFT becomes increasing inaccurate as the load on the helix becomes larger. This is because Lighthill's coefficients are designed to reproduce the SBT solution for a force-free helix. In this section we propose a set of load-dependent resistance coefficients which are designed to reproduce the SBT solution for a helix  {attached to a viscous load}, therefore correcting the inaccuracies resulting from using Lighthill's coefficients outside their regime of validity.
	
	\subsection{Derivation of optimal coefficients}

	Consider   a helix which 
	translates with velocity $U\zhat$ and rotates with angular velocity $\omega \zhat$. The result in Equation~\eqref{eq:rftmatrix} for the RFT solution may be rewritten as
	\begin{align}\label{eq:yeah}
		\begin{pmatrix}
			\alpha_\perp \\ \alpha_\parallel
		\end{pmatrix}
		= (1 + \epsilon^2)
		\begin{pmatrix}
			-h - \epsilon f  & \epsilon f - \epsilon^2 h \\\epsilon h + \epsilon^2 f & f - \epsilon h
		\end{pmatrix}^{-1	}
		\begin{pmatrix}
			a\omega \\U
		\end{pmatrix}.
	\end{align}
	We seek the `optimal' values of $\alpha_\perp$ and $\alpha_\perp$ such that for a given problem an RFT solution (Equation~\eqref{eq:rftmatrix}) would yield the same results for $(h,f)$ as an SBT solution (Equation~\eqref{eq:sbtmatrix}). If we use Equation~\eqref{eq:sbtmatrix} to eliminate $U$ and $a\omega$, we obtain that these optimal coefficients   satisfy
	\begin{align}\label{eq:coptimal}
		\begin{pmatrix}
			\alpha_\perp \\ \alpha_\parallel
		\end{pmatrix}
		= \frac{1 + \epsilon^2}{4\pi\eta}
		\begin{pmatrix}
			-h - \epsilon f  & \epsilon f - \epsilon^2 h \\\epsilon h + \epsilon^2 f & f - \epsilon h
		\end{pmatrix}^{-1	}
		\begin{pmatrix}
			-\Lambda_+&& \epsilon\Lambda_- \\
			-\epsilon\Lambda_- && \chi
		\end{pmatrix}
		\begin{pmatrix}
			h\\
			f
		\end{pmatrix},
	\end{align}
	where $h$ and $f$ are the SBT solutions for the problem of interest, i.e.~as given by (i) Equation~\eqref{eq:Uelim} for force-free swimming with angular velocity of flagellum prescribed; (ii) Equation~\eqref{eq:forcetorquefreematrix} for free swimming with angular velocity of flagellum relative to cell body prescribed; (iii) Equations~\eqref{eq:hmotortorque} and \eqref{eq:fmotortorque} for free swimming with torque applied by cell body onto flagellum prescribed. We find that all three cases yield the same optimal resistance coefficients, and we present here a general derivation which uses the force-free condition common to all three situations.
	
	The force balance condition Equation~\eqref{eq:forcebalance}
	may be used to eliminate $U$ in Equation~\eqref{eq:sbtmatrix},
	and the second row then yields a relation between $h$ and $f$,
	\begin{equation}\label{a2}
		f = \frac{\epsilon \lm }{\displaystyle \chi + \frac{4\pi \sqrt{a^2 + b^2} n}{3 R}}h.
	\end{equation}
	Substituting this in Equation~\eqref{eq:coptimal} and simplifying, we eliminate $h$ and $f$ and obtain 	\begin{equation}\label{eq:optimalsemiana}
		\alpha_\perp = \frac{\lp - \epsilon^2\lm + \rho(\lp\chi - \epsilon^2\lm^2)}{\displaystyle 4\pi\eta(1 + \rho(\chi + \epsilon^2\lm))}, \,\,\,\alpha_\parallel = \frac{\lp + \lm + \rho(\lp\chi - \epsilon^2\lm^2)}{\displaystyle 4\pi\eta(1 + \rho(\chi - \lm))},
	\end{equation}
	where $\rho := {3R}/({4\pi n \sqrt{a^2 + b^2})}$. 
	
	The result in Equation~\eqref{eq:optimalsemiana} is the 
	main result of our paper, namely optimal mobility coefficients which depend   on the geometrical parameters, the fluid viscosity, and (crucially) on the load via the dependence on $\rho$.  This is therefore a load-dependent RFT.

	\subsection{Analytical formulae for optimal coefficients}
	
	The expressions in Equation~\eqref{eq:optimalsemiana}, however, contain integrals which cannot be evaluated analytically. To obtain closed-form coefficients which may be readily used, we next expand rigorously the numerators and denominators to $\mathcal O(\epsilon^2,\epsilon_0^0,(1/\pi n)^0)$. We work to second order in $\epsilon$ because $\epsilon$ is generally largest (among $\epsilon,\epsilon_0$ and $1/\pi n$) in geometries of biological interest. Higher orders of $\epsilon$ produce very little gain in accuracy at the cost of considerably bulkier expressions. After lengthy algebraic manipulations, we  obtain  
	the results	
	\begin{equation}\label{coefffinal}
		\alpha_\perp = \frac{1}{4\pi\eta}\frac{P_1 + rP_2}{1 + rP_3},\,\,\,\,
		\alpha_\parallel = \frac{1}{4\pi\eta}\frac{P_4 + rP_2}{1 + rP_5},
	\end{equation}
	where
	\begin{align}\nonumber
		P_1 &= \lp^{(0)} + \epsilon^2(-\lm^{(0)} + \lp^{(2)}), \\\nonumber
		P_2 &= \chi^{(0)}\lp^{(0)} + \epsilon^2\left[-(\lm^{(0)})^2 + \chi^{(2)}\lp^{(0)} + \chi^{(0)}\left(\frac{\lp^{(0)}}{2} + \lp^{(2)}\right) \right], \\\nonumber
		P_3 &= \chi^{(0)} + \epsilon^2\left(\frac{\chi^{(0)}}{2} + \chi^{(2)} + \lm^{(0)}\right), \\\nonumber
		P_4 &= \lm^{(0)} + \lp^{(0)} + \epsilon^2(\lm^{(2)}+\lp^{(2)}), \\
		P_5 &= \chi^{(0)} - \lm^{(0)} + \epsilon^2\left[\chi^{(2)} + \frac{\chi^{(0)}-\lm^{(0)}}{2} - \lm^{(2)}\right],\,\,
		r = \frac{3R}{4\pi n b},
	\end{align}
	and 
	\begin{align}\nonumber
		\Lambda_\pm^{(0)} &= - \ln\epsilon_0 \pm \frac{1}{2} - \gamma + \ln 2 ,\,\,\,\,
		&&\Lambda_\pm^{(2)} = \mp \ln\epsilon_0 - \frac{1}{4} \mp \gamma + \ln 2, \\
		\chi^{(0)} &=  - 2\ln\epsilon_0 + 2\ln(\pi n) -1 + 2 \ln 2,\,\,\,\,
		&&\chi^{(2)} = \ln \epsilon_0 + \ln (\pi n) -\frac{1}{2} + 2\gamma - \ln 2, 
	\end{align}
	and $\gamma \approx 0.5772$ is the Euler-Mascheroni constant.
	
	Since this calculation did not involve the rotational drag coefficient of the load, our results hold for a load of a general shape, with $R$ interpreted as the radius of a sphere with the same drag coefficient as the load in  question. This calculation, however, relies crucially on the force balance equation in the $z-$direction (i.e.~force-free motion, Equation~\eqref{eq:forcebalance}). While this does hold for a wide range of problems of interest in biology, our coefficients would not be valid, for instance, for a helix sedimenting under gravity.

	\subsection{Implications on flow field and swimming speed}
	Equipped with our new load-dependent RFT, we now revisit the problem of the flow field induced by a helix rotating without translation (\S5(b)\ref{sec:flowfieldinhelix}) and the free swimming speed of a helical swimmer with a load (\S5(b)\ref{sec:swimmingspeeds}), this time using our optimal resistance coefficients expanded as above to second order in $\epsilon$. 
	
	We plot in Fig.~\ref{fig:flowinhelix}c the $(x,y)$-component and in Fig.~\ref{fig:flowinhelix}f the $z$-component of the flow in the $z = 0$ plane by the helix rotating about its axis. Using our optimal coefficients (taking the limit $r \to \infty$) over Lighthill's coefficients further reduces the already small errors in the $(x,y)$-components, and, crucially, shows a marked improvement in the accuracy of the $z$-component of flow. 	Similarly, using RFT with these optimal load-dependent coefficients to solve the swimming problem produces much more accurate results than Lighthill's coefficients. We next plot the swimming speeds as obtained by these optimal coefficients in Fig.~\ref{fig:swimmingspeeds}, in blue, and see much closer agreement with numerics.

	We further quantify the deviation of RFT results from the numerical solution by defining an error using the difference in the areas under the curves, $E := |\int_0^{0.3L} [U^{\text{SBT}}(R) - U^{\text{RFT}}(R)] \,\mathrm dR|/|\int_0^{0.3L} U^{\text{SBT}}(R) \,\mathrm dR|$.	From the speed-load radius curves for the normal polymorph illustrated in Fig.~\ref{fig:swimmingspeeds}a, we obtain values of the errors incurred by using Lighthill's RFT  {and} our newly-derived load-dependent RFT,  {respectively}. We 
	repeat this computation for the geometrical parameters corresponding to the other flagellar polymorphic forms, {see Table~\ref{table}.}  We then plot these errors for each polymorphic form in Fig.~\ref{fig:swimmingspeeds}b. In all cases,  the load-dependent coefficients consistently outperforms Lighthill's coefficients, often by orders of magnitude.

	\begin{figure}[t]
		\centering	\includegraphics[width=0.7\textwidth]{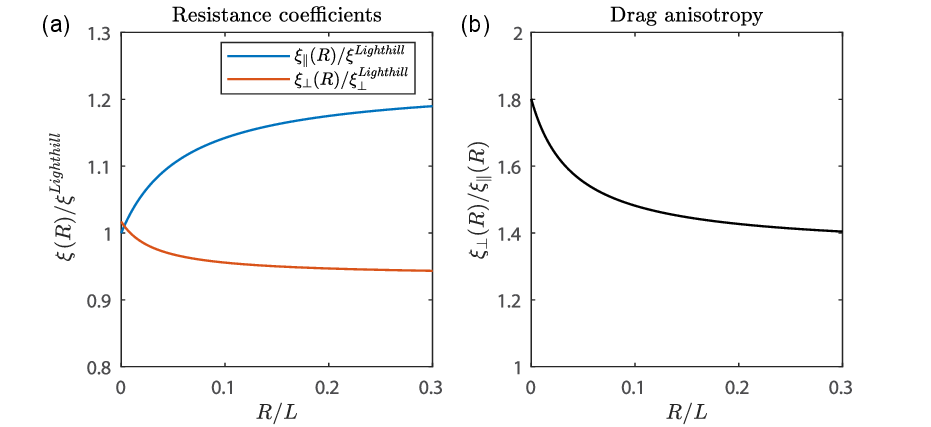}
		\caption{{\bf Optimal resistance coefficients.} Values of $\xi_\parallel(R)$ (blue) and $\xi_\perp(R)$ (red) 
			relative to Lighthill's coefficients 
			$\xi_\parallel^{Lighthill}$ and $\xi_\perp^{Lighthill}$ 
			respectively, plotted against $R/L$. (b) Drag anisotropy 
			$\xi_\perp(R)/\xi_\parallel(R)$ against $R/L$. Geometrical 
			parameters $\epsilon = 0.61, \epsilon_0 = 0.028, n = 3.8$ 
			corresponding to a normal polymorph are 
			used. 
		}\label{fig:coeffs}
	\end{figure}

	We next show in Fig.~\ref{fig:coeffs} how these load-dependent coefficients vary with the load, i.e.~the effective hydrodynamic radius of the load $R$. 	We plot these coefficients in their more common resistance forms, $\xi_{\parallel,\perp} = 1/\alpha_{\parallel,\perp}$. 
	The helix is assumed to have the geometrical parameters of the normal polymorph. The parallel resistance coefficient increases by up to 20\% as $R$ increases, whereas the normal resistance coefficient decreases but by a much smaller proportion;  
	this is expected since the load-dependence enters the normal coefficient only at order $\epsilon^2$.
	We will provide a physical explanation for the observed load-dependence in the next section. 	We further plot the drag anisotropy $\xi_\perp/\xi_\parallel=\alpha_\parallel/\alpha_\perp$ against $R$ (Fig.~\ref{fig:coeffs}) and see that its value decreases with $R$; the decrease in swimming speeds with load therefore stems not only directly from the viscous resistance of the load, but also from a decrease, through the load-dependent coefficients, in the drag anisotropy necessary for flagellar propulsion~\cite{lauga2009, hancock1953, becker2003}.

	Finally, it is instructive to consider the leading order expansions in the limit $\epsilon \ll 1$  of the optimal coefficients 
	\begin{equation}
		\alpha_\perp^{(0)} = \frac{\lp^{(0)}}{4\pi\eta},\quad
		\alpha_\parallel = \frac{\Lambda_+^{(0)} + \Lambda_-^{(0)} + r\Lambda^{(0)}_+\chi^{(0)}}{4\pi\eta[1 + r(\chi^{(0)} - \Lambda_-^{(0)})]}.
	\end{equation}
	
	Rewriting 
	\begin{equation}
		\Lambda_+^{(0)} = \log \left(\frac{\ell}{r_0}\frac{e^{-\gamma}}{\pi}\right) + \frac{1}{2}, \quad \Lambda_+^{(0)} + \Lambda_-^{(0)} = 2{\log \left(\frac{\ell}{r_0}\frac{e^{-\gamma}}{\pi}\right)},
	\end{equation}
	where $\ell = 2\pi b$ in the $\epsilon \ll 1$ limit, we observe that we recover Lighthill's coefficients at $r = 0$ since $e^{-\gamma}/{\pi} \approx 0.18$.  {This also emphasises that the normal mobility coefficient $\alpha_\perp$ is load-independent at this order. Noting that  $h = -a\omega/\alpha_\perp$ at leading order (as can be shown from Equation~\eqref{eq:rftmatrix}) and that the dominant contribution to the flow $(u,v)$ induced by a helix rotating without translation depends only on $h$ (Equations~\eqref{eq:asym1} and \eqref{eq:asym2}), we deduce that the dominant contribution to $(u,v)$ depends only on $\alpha_\perp$. This explains on a more rigorous footing why Lighthill's coefficients in the case of a helix rotating without translating accurately reproduce the $(x,y)$-components of the flow field.}

	\subsection{Physical interpretation of load-dependence}
	
	The effective load-dependence of the optimal resistance coefficient may, a priori, be  a surprising result, since the helix and the load do not interact hydrodynamically in our setup, except through a viscous drag. What is the  physical origin of this dependence? To answer this we consider the two limiting cases: force-free swimming ($R = 0$) and no translation ($R \to \infty $). We   work to leading order in $\epsilon$, which will be sufficient to reveal the  physical origin of our result.

	Consider a  right-handed helical filament rotating at an angular velocity $\omega \zhat$ and undergoing force-free motion (i.e.~$f = 0$). Using SBT (i.e.~Equation~\eqref{eq:sbtmatrix}), the radial component $h$ of the force distribution associated with this motion, and the resultant swimming speed $U_{\text{free}}$, are given by
	\begin{equation}\label{eq:hUsbt}
		h^{SBT}_{\text{free}} = -\frac{4\pi\eta a\omega}{\lp}, \,\,\,\,\,\,\,U_{\text{free}}^{SBT} = -\frac{\lm \epsilon h}{4\pi\eta}.
	\end{equation}
	In contrast,	RFT gives, by Equation~\eqref{eq:rftmatrix},
	\begin{equation}\label{eq:hUrft}
		h^{RFT} = -\frac{a\omega}{\alpha_{\perp}},\,\,\,\,\,\,\, U_{\text{free}}^{RFT} = -(\alpha_\parallel - \alpha_\perp)\epsilon h.
	\end{equation}
	Lighthill's coefficients $\alpha_\parallel = 2\pi\eta/(\log 2\eninv - \gamma)$ and $\alpha_\perp = 4\pi\eta/(\log 2\eninv - \gamma + 1/2)$ are chosen such that RFT gives the same predictions for $h$ and $U_{\text{free}}$ as SBT i.e.~$h^{SBT} = h^{RFT}$ and $U_{\text{free}}^{SBT} = U_{\text{free}}^{RFT}$.

	Crucially, the force distribution along the helix has no axial component i.e.~$f = 0$, for both RFT and (our helically symmetric approximation of) SBT. This means that the Stokeslets along the helix are all oriented radially and thus the effects of Stokeslets in the far field sum up to approximately zero; this has the effect of localising hydrodynamic interactions. The $\log 2\eninv - \gamma$ factor in Lighthill's coefficients reflect the fact that a point on the helix effectively only experiences hydrodynamic interactions from within a distance of $q = b/e^{\gamma}$, or equivalently, 0.09 times the helical wavelength.

	Consider now the same helix rotating but prevented from translating. In the following we take $\alpha_\perp$ and $\alpha_\parallel$ to be the Lighthill coefficients, since we wish to compare SBT with Lighthill's RFT. The force distribution associated with this motion necessarily has a non-zero $f$. 
	
	We first show that $f$ must have a smaller magnitude under SBT than under Lighthill's RFT. Using SBT (Equation~\eqref{eq:sbtmatrix} with $U=0$), we obtain
	\begin{equation}
		h^{SBT} = -\frac{4\pi\eta a\omega}{\lp}, \,\,\,\,\,\,\,f^{SBT} = -\frac{\epsilon \lm h}{\chi},
	\end{equation}
	while RFT yields
	\begin{equation}
		h^{RFT} = -\frac{a\omega}{\alpha_{\perp}},\,\,\,\,\,\,\, f^{RFT} = -\frac{\epsilon (\alpha_\parallel - \alpha_\perp) h}{\alpha_\parallel}.
	\end{equation}
	We may rewrite these expressions for $f$ as
	\begin{equation}\label{eq:fphy}
		{\alpha_\parallel}f^{RFT} = -U_{\text{free}},\,\,\,\,\,\,\,\frac{\chi}{4\pi\eta}f^{SBT} = -U_{\text{free}}.
	\end{equation}
	Here $U_{\text{free}}$ is the force-free swimming speed calculated earlier; this takes the same value $U_{\text{free}} = U_{\text{free}}^{SBT} = U_{\text{free}}^{RFT}$ under SBT and RFT since we are using Lighthill's coefficients. Physically, at each point on the helix, $f$ balances the local axial velocity $U_{\text{free}}$ at which the filament would move if it were force-free, thus ensuring that the helix does not translate.

	Under RFT, it is only the local $f$ which balances $U_{\text{free}}$. Lighthill's coefficient $\alpha_\parallel$ derived for force-free motion does account for hydrodynamic interactions, but only locally, up to a distance $q$ away, as discussed earlier. In contrast, under SBT, the local $f$, as well as $f$'s from distant sections of the helix, all contribute towards balancing $U_{\text{free}}$. Because of this cooperation, a smaller value of $f$, is required to balance $U_{\text{free}}$ compared to RFT. The inclusion of long-range hydrodynamic interactions from the entire helix is reflected in the $\log \pi n$ term in $\chi$.

	We have thus established why $f$ in the no-translation case has a smaller magnitude under SBT than under Lighthill's RFT. The optimal tangential resistance coefficient in the no-translation case may be written as (working from Equation~\eqref{eq:yeah})
	\begin{equation}\label{yes}
		\xi_\parallel = \frac{f - \epsilon h}{\epsilon a \omega} = \frac{\epsilon|h| - |f|}{\epsilon a \omega},
	\end{equation}
	where the second equality used that both $h, f < 0$ when $\omega > 0$. The signs of $h$ and $f$ reflect the directions of the external forces on the helix required to sustain an $\omega > 0$ rotation and balance out the tendency of the helix to swim in the $+\zhat$ direction. We know that 
	$h$ takes the same value under SBT and Lighthill's RFT, whereas $f$ takes a smaller value under SBT. Therefore the optimal $\xi_\parallel$ in the no-translation ($R \to \infty$) case is {larger} than Lighthill's $\xi_\parallel$. In the force-free ($R = 0$) case, however, the optimal $\xi_\parallel$ \textit{is} Lighthill's coefficient. As $R$ changes, the optimal $\xi_\parallel$ must vary smoothly between the $R = 0$ value and the $R\to\infty$ value.

	The physical origin of the load-dependence of the optimal tangential resistance coefficient is therefore the long-range hydrodynamic interactions between axial point forces.  
	

	\section{Conclusion}
	
	A careful examination of the assumptions made in Lighthill's derivation of a resistive-force theory for helical filaments reveals a limitation largely overlooked in the literature: the  resistance coefficients were derived under the specific conditions of a force-free helix rotating with no external load attached.  These conditions do not hold for most biological or bioinspired systems of interest, which involve helical filaments attached to some external load, {{but Lighthill's coefficients are nonetheless often used to model these systems. In this manuscript}}    we have first illustrated, using the flows created by a helix rotating without translation and the swimming speeds of a helical swimmer, the inaccuracies resulting from using Lighthill's coefficients when a load is present. We have then revisited Lighthill's problem rigorously in the presence of an external load attached to the helix, using both RFT and SBT. By, in essence, matching RFT results with SBT results, we have derived optimal resistance coefficients which reproduce SBT results, thus offering a significant improvement upon Lighthill's coefficients. These optimal coefficients involve integrals which one must evaluate numerically, but we have shown that asymptotic expansions of these coefficients (to second order in $\epsilon$) yields analytical expressions with very little compromise  on accuracy.  Surprisingly, our optimal coefficients depend on the size of the load, which we have interpreted  physically as a consequence of the  long-range hydrodynamic interactions between axial point forces along the helix. 
	
	{Although we have accounted for the load through the viscous drag in the force balance, we have made the  modelling choice of neglecting  direct hydrodynamic interactions between the sphere and the helix. This allowed us to exploit helical symmetry and derive analytical expressions for the optimal resistance coefficients. In order to analytically estimate the error incurred by this assumption, we may carry out a simplified  calculation for a sphere-rod swimmer driven by a slip velocity along the rod, which acts as a proxy for the full helical swimmer system. We can calculate the resultant swimming speed {in the case where direct hydrodynamic interactions between the sphere and rod are accounted for (via Stokeslets distributed along the rod and at the sphere centre), and in the case where the sphere and rod interact only via the force balance.} Comparing these results, we find that neglecting these hydrodynamic interactions overestimates the swimming speed by less than 10\% for the geometrical parameters of an \textit{E.~coli} normal polymorph. }
	
	{This paper represents an initial step towards incorporating the effects of a load into a resistive-force theory. Future studies may explore the inclusion of hydrodynamic interactions to enhance quantitative accuracy. Extending our approach to different filament geometries presents an intriguing challenge, and we anticipate that the load dependence demonstrated here will remain {broadly valid.}}
	

	\section*{Acknowledgements}
	
	{Financial support from the Cambridge Trust (scholarship to P.H.H.) is gratefully acknowledged.}
	

		\section*{Data Availability}
		Supporting code is available in the Zenodo repository \url{https://zenodo.org/records/15060067} .

	\bibliography{LDRFT_references.bib}
		
	\end{document}